# Temperature dependence of nitrogen-vacancy center ensembles in diamond based on an optical fiber

Ke-Chen Ouyang[1,2], Zheng Wang[2,3], Li Xing[2], Xiao-Juan Feng[2,*], Jin-Tao Zhang[2], Cheng Ren[1], Xing-Tuan Yang[1]

[1]*Institute of Nuclear and New Energy Technology, Tsinghua University, Beijing 100084, China*

[2]*National Institute of Metrology, Beijing 100029, China*

[3]*Department of Precision Instrument, Tsinghua University, Beijing 100084, China*

The nitrogen-vacancy (NV) centers in diamond sensing has been considered to be a promising micro-nano scale thermometer due to its high stability, good temperature resolution and integration. In this work, we fabricated the sensing core by attaching a diamond plate containing NV centers to the section of a cut-off multi-mode fiber. Then we measured the zero-field splitting parameter ($D$) of NV center ensembles using continuous-wave optical detected magnetic resonance (CW-ODMR) technique. A home-made thermostatic system and two calibrated platinum resistance thermometers were applied for reference temperature measurement. The effects from preparation time and count time in the pulse sequence, laser power, microwave power, and microwave frequency step were investigated. Moreover, the experimental $D$ and $T$ from 298.15 K to 383.15 K was obtained with the standard uncertainty of $u(D)$ = (3.62268~8.54464) × $10^{-5}$ GHz and $u(T)$ = (0.013~ 0.311) K. The experimental results are well consistent with the work of Toyli, et al. (Toyli, et al., 2012) using the similar diamond sample. The extrapolation for $D$-$T$ at 0 K and 700 K also agree with other references, and meanwhile $\mathrm{d}D/\mathrm{d}T$ varies with temperature. Finally, comparing the $D$-$T$ relationship measured by different research groups, we can know that the NV concentration resulting in different electron density and manufacturing procedure resulting in different thermal expansion would lead to different $D$-$T$ relationship. It is worthy to continue further comprehensive research especially from the metrological point of view to develop NV center as a practical and accurate micro-nano scale thermometry.

## I. INTRODUCTION

The nitrogen-vacancy (NV) center in diamond has been applied to detecting several physical quantity including magnetic field [1,2], electric field [3,4], pressure [5,6], and temperature [7-9] with high accuracy and short response time. The diamond thermometer has been applied for nano-scale temperature measurement, such as in cells [10,11], in organisms [12,13], on chips [14,15], et al. The electron spin of NV centers can be polarized by laser and the spin state can be manipulated by microwave of certain frequency. Thus, the NV center have ultra-high sensitivity to measure temperature by the variety of zero-field splitting energy ($D$) [8,10,15]. Due to the nano-scale of NV centers and the physical stability of diamond, the thermometry has advantages on high resolution and robustness [7,16]. The active control of laser polarization and microwave manipulation enables it to be resistive to ambient environmental by encoding certain sequences of laser and microwave [17,18]. Recently, the technology for optical fiber probe coupled with NV centers makes it have the potential of integration [19-21,15]. High-resolution temperature measurement is based on NV centers with an accurate calibration of $D$-$T$ relationship. However, the parameters of reference temperature, laser, microwave and diamond sample have influence on the acquisition of the $D$-$T$ relationship. Therefore, it is significant to improve the reliability of calibrating the $D$-$T$ relation.

* fengxj@nim.ac.cn

Several groups have researched on the relationship between $D$ and $T$. According to Ref. [22], the temperature dependence of NV centers magnetic resonance from 280 K~330 K was studied, and the relation of $\mathrm{d}D/\mathrm{d}T$ = -74.2(7) kHz/K was got in 2010. Besides, the group used different diamond samples for experimental and obtained different values of $\mathrm{d}D/\mathrm{d}T$. A fifth-order polynomial relation of $D$ and $T$ from 5.6 K to 295 K and a third-order polynomial relation from room temperature to 700 K were obtained in 2011 and 2012, respectively [23,14]. In 2014, another second-order polynomial relation of $D$-$T$ from 300 K to 600 K was worked out and the result was found in good consistency with that in Ref. [9,14]. In 2018, a similar rate of frequency shift over temperature change with that in Ref. 22, which is equal to -74.2 kHz/K, was acquired from 297 K to 333 K [24]. In conclusion, present $D$-$T$ relations vary with different groups and it is necessary to carry on careful experiments to provide reference for the application of NV centers thermometry.

In this work, to obtain a good thermal equilibrium between the NV center and reference thermometers, we fabricated a sensor based on optical fiber and calibrated it in a close thermostatic cavity. We attached the diamond sample with NV centers to the section of a cut-off multi-mode optical fiber and built up a continuous-wave optical detected magnetic resonance (CW-ODMR) system and a home-made thermostatic system. Then we determined the optimal combination of ODMR experimental parameters by several groups of tests, based on which we carried out the ODMR experiments from 298.15 K to 383.15 K with a step of 5 K. The corresponding zero-field splitting parameter($D$) in every temperature value($T$) was obtained by fitting the spectra data with the most appropriate function. Accordingly, we attained the relationship between $D$ and $T$ in the temperature range from 298.15 K to 383.15 K and analyzed uncertainty budget sources in detail. Finally, we compared the results with those of other research groups to demonstrate the consistency and discussed the connection between different samples and $D$-$T$ relationship.

# II. EXPERIMENTAL SETUP AND MEASUREMENT PROCEDURE

## A. Principle of temperature measurement by NV centers

The physical quantity that can be detected by the NV center is determined by the Hamiltonian, which is expressed as [8]

$$H_{\mathrm{NV}} = DS_z^2 + g_s\mu_B \boldsymbol{B}\cdot\boldsymbol{S} + \boldsymbol{S}\bar{A}\boldsymbol{I} \tag{1}$$

where $D$ is the zero-field splitting parameter between the sublevels $m_s$=0 and $m_s$=±1. $\boldsymbol{S}$ represents spin angular momentum with its component $S_x$, $S_y$, $S_z$ in three axis, respectively. $g_s$ is dimensionless magnetic moment (Landé $g$-factor) and $\mu_B$ is Bohr magneton. $\boldsymbol{B}$ stands for vector external magnetic field. $\bar{A}$ is the hyperfine tensor and $\boldsymbol{I}$ is the spin operation of nitrogen nucleus. The first item of of the right-side of the equation (1) reflects the influence from temperature, the second item reflects the influence by magnetic field on the fine structure of NV center, while the third one reflects the influence by strain and electric field.

The influence by temperature on NV center is estimated by the variation of the zero-field splitting parameter $D$. The ground state of NV center is spin-triplet with the sublevels $m_s$ = 0 and $m_s$ =±1. The sub-levels $m_s$ = ±1 are degenerated when there is no external magnetic field applied. Between the sublevels $m_s$ = 0 and $m_s$ =±1 exists a splitting called the zero-field splitting, which can be represented by the parameter $D$. The value of $D$ changes with the temperature of the spot where the NV center locates. Thus, we can measure the value of $D$ by ODMR technique and then obtain the corresponding temperature $T$ by calibrating the relationship between $D$ and $T$.

According to the $NV^-$ molecular model, the expression of $D$ can be yielded as [25]

$$D \approx C\eta^2 \langle \frac{1}{r_{12}^3} - \frac{3z_{12}^2}{r_{12}^5} \rangle \qquad (2)$$

where $C$ is the spin-spin interaction constant, $\vec{r}_i = x_i\hat{x} + y_i\hat{y} + z_i\hat{z}$ is the position of the $i$th electron, of which $r_{12} = |\vec{r}_2 - \vec{r}_1|$, $z_{12} = z_2 - z_1$, $\eta$ is the electron density. The value of $D$ is dependent of $\eta$ and $\langle \frac{1}{r_{12}^3} - \frac{3z_{12}^2}{r_{12}^5} \rangle$, which represents the interaction between the dangling $sp^3$ electron densities of the two carbon atoms. Rise in temperature will cause local thermal expansion in diamond lattice, and then increase the value of $r_{12}$. Thus, the value of $D$ will correspondingly decrease with $\eta$ unchanged and the variety of $D$ can be observed by measuring the ODMR spectrum.

## B. Experimental setup

The experimental setup is based on CW-ODMR technique and an optical fiber thermometer coupled with NV center ensembles to realize highly reliable temperature measurement. The experimental system, as showed in Fig. 1, includes five parts: the laser polarization system, the sensing core, the microwave manipulation system, the fluorescence detecting system, and the thermostatic system. In the ODMR experiment, a polarization beam at 532 nm is emitted by a laser (CNI-MGL-III-532-200mW) and modulated by an acousto-optic modulator (AOM, AODR 1080AF-DIF0-1.0 from Gooch & Housego), which can polarize the electron spins of NV centers. The microwave manipulation system is applied to provide sufficiently high power microwave for NV centers to change the spin states, in which an amplifier (ZHL-16W-43-S+ from Mini-Circuits) is set to amplify the microwave from the microwave generator (SMIQ 06B from Rohde & Schwarz) and a RF switch (ZASWA-2-50DRA+ from Mini-Circuits) controls the on-off of the microwave signal. Fluorescence released by the polarized NV centers can be detected by a single photon count modulator (SPCM, SPCM-AQRH-11-FC from Excelitas). Then the data is collected by a data acquisition (DAQ, USB6363 from National Instrument) connected with the SPCM. To better control the laser, the microwave and the readout to switch on or off, the AOM, the RF Switch and the DAQ are all connected with a pulse generator (PBESR-PRO-500-PCI from SpinCore), which is inserted in the computer. Finally, the CW-ODMR spectra can be measured by frequency sweeping.

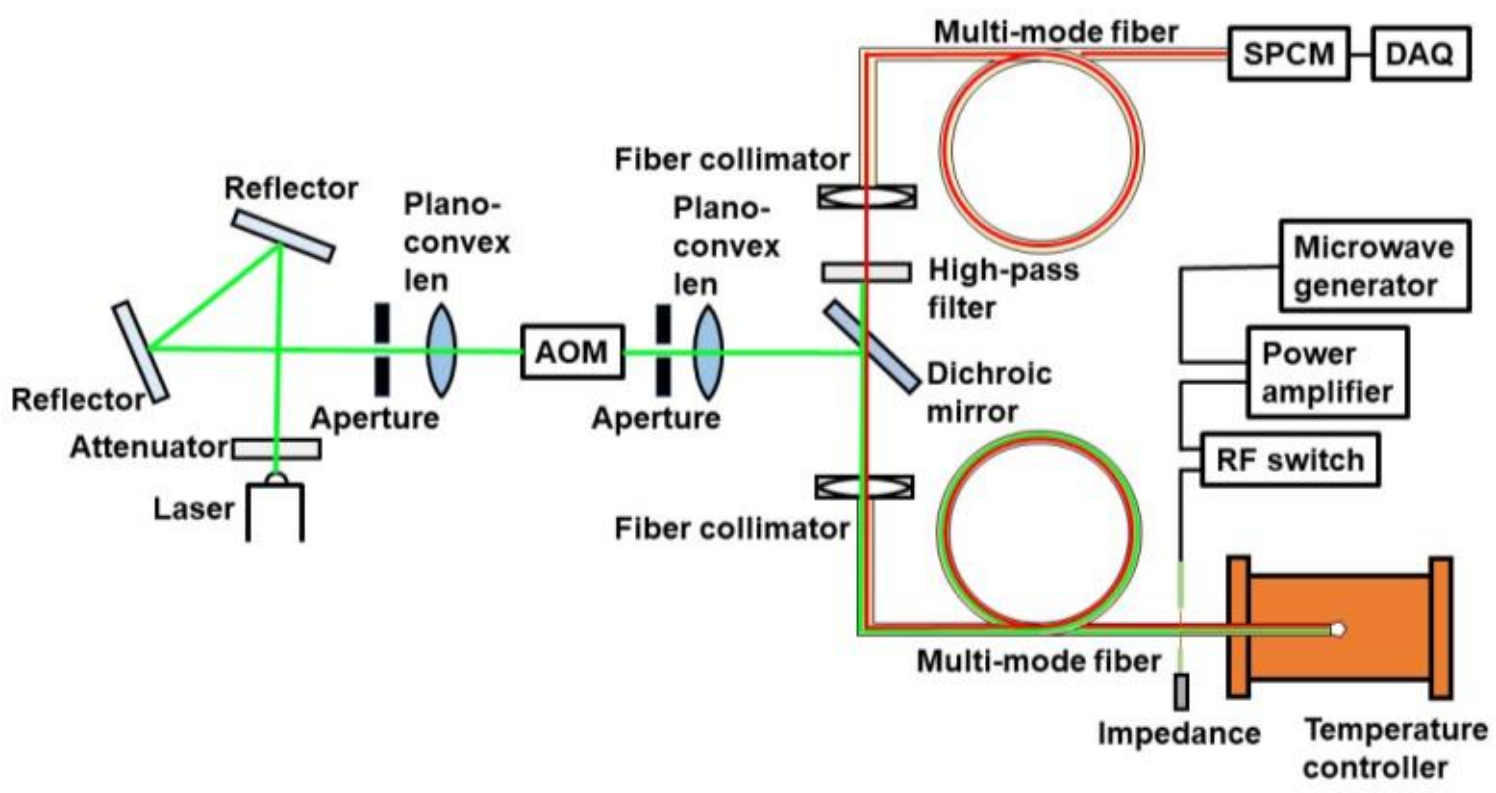

FIG. 1. The schematic diagram of the experimental system. AOM: acousto-optic modulator. SPCM: single photon count modulator. DAQ: data acquisition.

The sensing core consists of a multi-mode optical fiber, a diamond plate (Element Six EL SC Plate, 2.0 mm × 2.0 mm × 0.5 mm, [$Ns^0$] <5 ppb) and a copper wire (60 μm in diameter), as shown in Fig. 2. The fiber is 1 m in length and we cut it in half with a ceramic fiber scribe. Then we attached the diamond plate to the section of the fiber with UV-curing optical adhesives. Next, we stuck the copper wire to the plate using the same glue and bond it to the fiber with adhesive tape and raw material belt. The wire has two free ends, one of which is connected with a high power amplifier through a coaxial cable while the other is connected with a resistance by another cable to absorb the microwave.

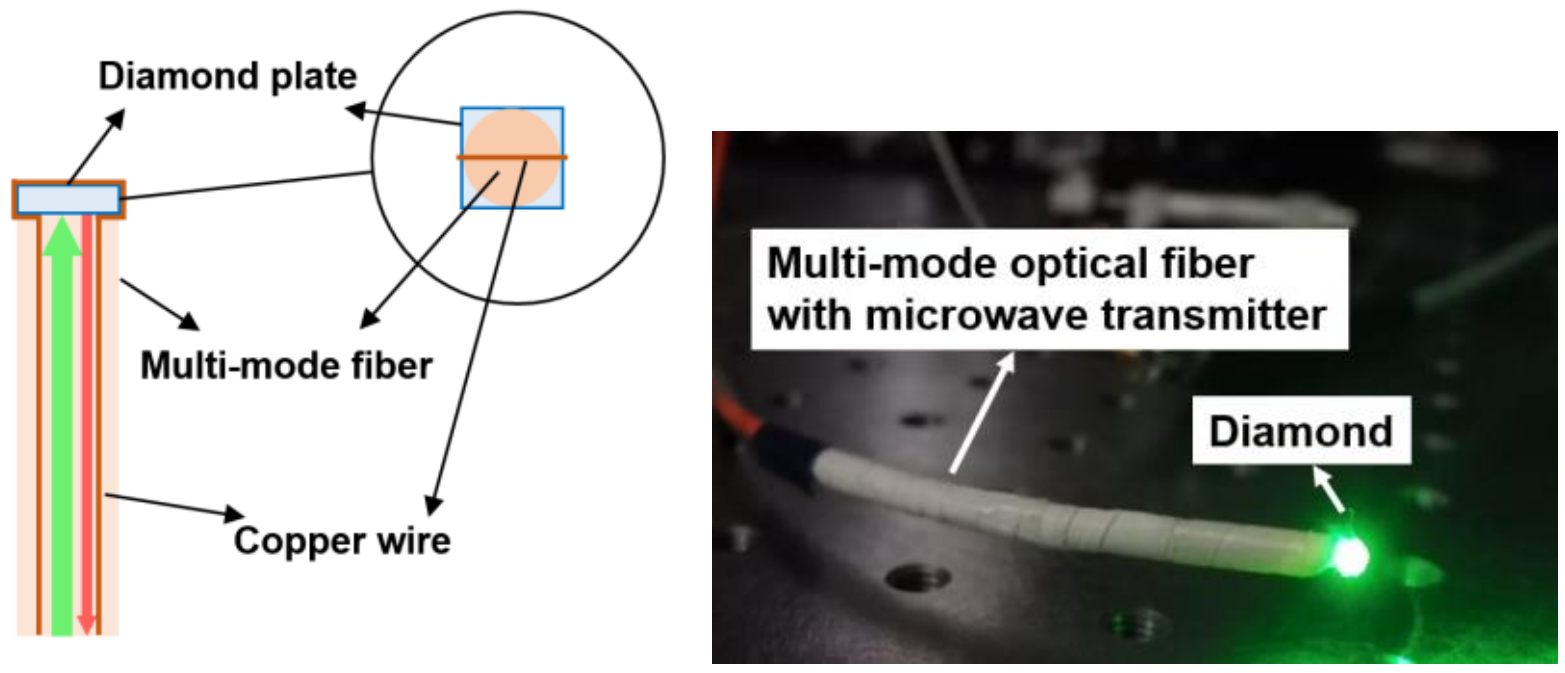

FIG. 2. The schematic diagram and the picture of the sensor.

To form a temperature field with high stability and homogeneity, we set up a close thermostatic cavity, which is comprised of a self-designed temperature controller, a heater, a DC power supply, a millikelvin thermometer (MKT 50 from Anton Paar), two platinum resistance thermometers (Pt-100) and a PC. The schematic diagram is illustrated in Fig. 3. The main body of the system is the temperature controller, which includes a cylindrical chamber, a copper block, a front end cap and a back end cap. The block is cylindrical with screw thread on the side. There are three straight holes in the block. Two openings of the holes are on the same underside of the block, which are equal in length and for the Pt-100. The other one is on the other underside of the block and is designed for the fiber. Seen from the cross section of the block, the one hole for the fiber is in the middle, while the two holes for the Pt-100s are located on the circle centered on the fiber hole. There is an overlap of the holes in the axial direction, which aims to make the sensors of the fiber (the diamond) and the Pt-100 in the same position on the axis. When the block is matched into the chamber, the underside of the block with the fiber hole should be on the same side with the front end cap, which also has a fiber hole in the middle. The back end cap with a screw hole for a Lemo connector, which converges the leads of the Pt-100 and connects them with the thermometer, is installed on the same side with the other underside of the block. A heating film is attached around the chamber and in the outermost cotton wraps with insulation cotton. The PC controls the DC power supply to supply power to the heater and reads the temperature data from the MKT 50. These devices constitute a PID control system, which functions to keep the temperature controller at the set temperature. Before the experiments, two platinum resistance thermometers were calibrated in National Institute of Metrology, China in a stable liquid thermostatic bath with standard platinum resistance thermometers as reference. The calibration uncertainty is less than 0.007 K.

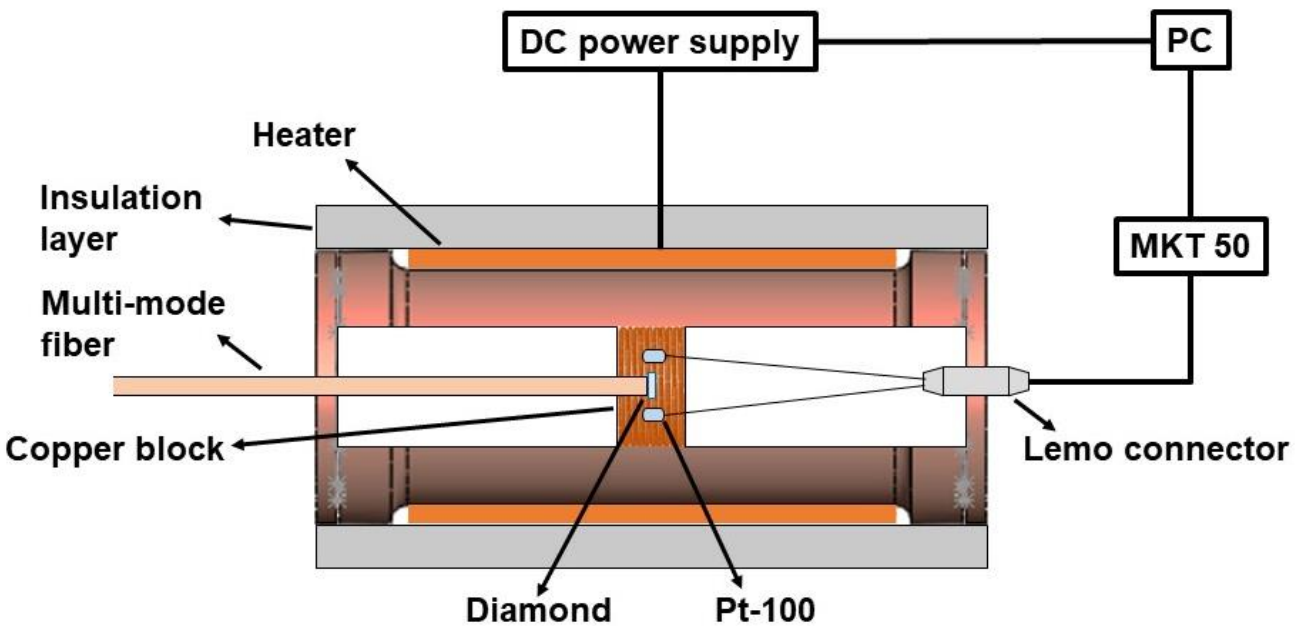

FIG. 3. The schematic diagram of the thermostatic system.

# III. EXPERIMENTAL RESULTS AND DISCUSSION

## A. ODMR measurement results

In the CW-ODMR experiment, the pulse generator is coded with a certain sequence (Fig. 4) to add an extra count whose duration is as long as the count time at each frequency point. The purpose of setting the preparation time is to ensure that the supply of microwave is stable when reading data and count time is the duration during which the data acquisition is reading data at every microwave frequency point. The data acquired within the extra count is regarded as the reference and the normalized intensity of fluorescence can be output directly. The aim of introducing the reference is to reduce the impact of laser fluctuations and keep the performance of data at a relatively same level.

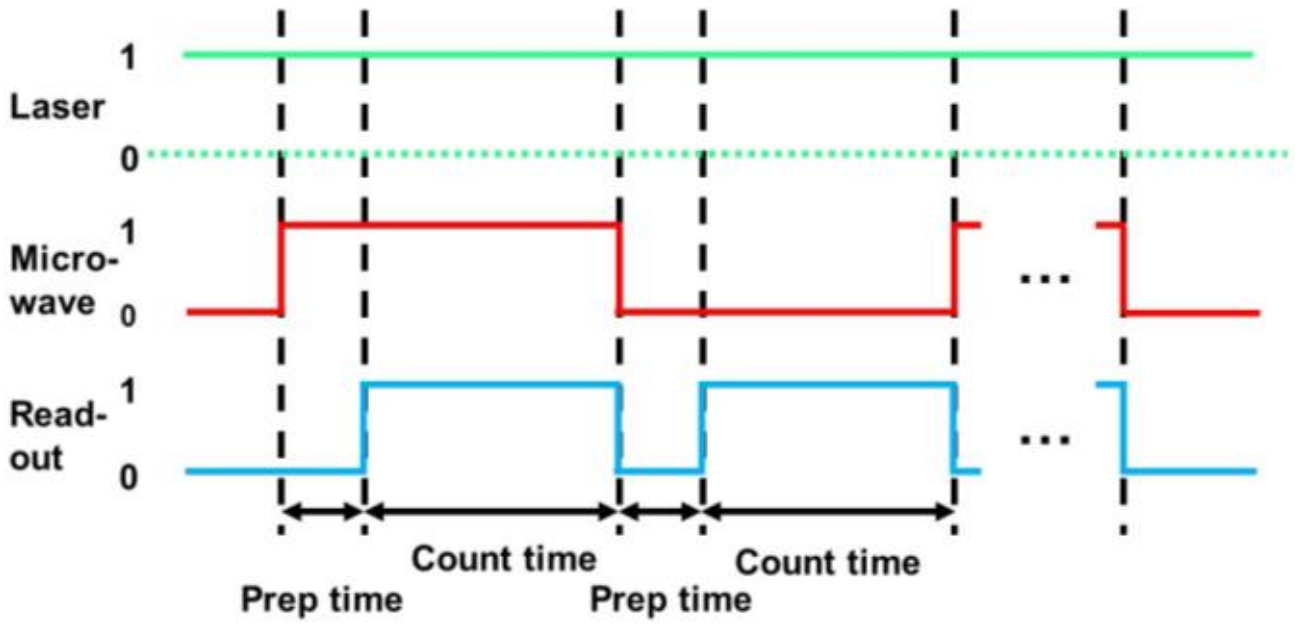

FIG. 4. The schematic diagram of the pulse sequence.

A typical ODMR spectrum at 303.15 K is shown in Fig. 5. In principle, the sublevels $m_s = \pm 1$ of the ground state are degenerated when no external magnetic field is applied. Therefore, there ought to be one dip in the spectrum. However, we observed double-peak structure emerging in the spectrum during the experiment even when there was no external magnetic field, as shown in Fig. 5(a). Such a phenomenon can be attributed to local strain, which removes the degeneracy of the ground sublevels $m_s = \pm 1$ [21]. Researchers believed that the ODMR line conforms to the Lorentzian line shape and used it to fit the ODMR [26-28]. So we used Lorentzian function to do fitting as previous papers, as well as another approximative function, Gaussian function in early researches. However, these two functions were more suitable to fit single-peak structure and could not obtain high fitting accuracy for double-peak structure. Finally, we use InvsPoly function as a new fitting equation to improve the accuracy, as follows,

$$y = y_0 + \frac{A}{1+A_1[2(x-x_c)/w]^2+A_2[2(x-x_c)/w]^4+A_3[2(x-x_c)/w]^6} \quad (3)$$

where $A$, $A_1$, $A_2$ are the fitting coefficients and $x_c$ is the middle frequency of the double-peak, that is the $D$. It has been demonstrated that the function could fit double-peak structure more effectively to represent local strain.

The comparison on fittings with Lorentzian function, Gaussian function and InvsPoly function were also showed in Fig.5. The residuals of Fig. 5(a) shows that the InvsPoly curve is random and is in closer proximity to the shape of the spectrum. Additionally, the InvsPoly fitting leads to a substantial increase in $R$-square and decrease in error compared with the other two as shown in Fig. 5(b). Consequently, we selected the InvsPoly function as the fitting function in later researches.

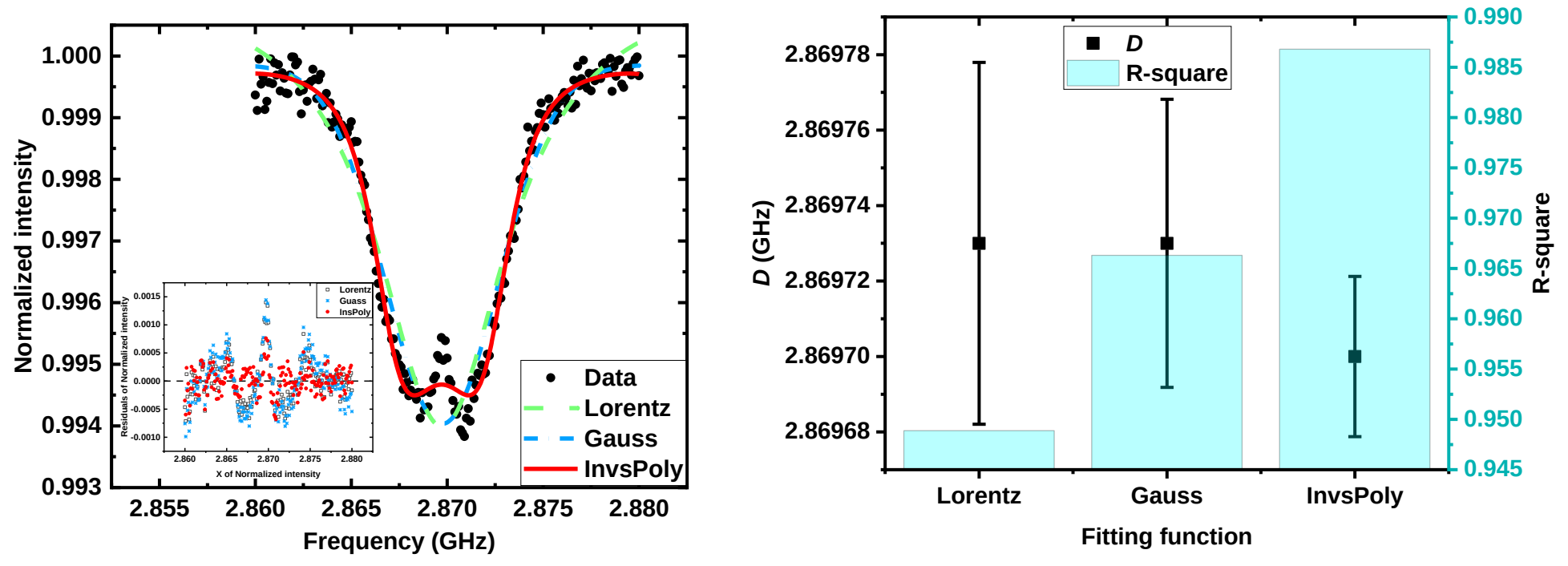

(a) (b)

FIG. 5. ODMR results. (a) The graphs of the fitting results with Lorentzian function, Gaussian function and InvsPoly function. The inserted picture is the distribution of residuals of the fitting by three functions. (b) $D$, $\delta D$ and $R$-square obtained with three functions.

## B. Effects of the measurement parameters

The experiment parameters, including the preparation time and the count time of the pulse sequence, the output power (current) of the laser, the output power of the microwave generator, the frequency step of the microwave and the count time, have a remarkable effect on the stability and accuracy of the experiment. To figure out relatively more precise value of $D$, the optimal combination of these parameters should be determined. We recognized that these parameters were independent with each other. We changed a single variable while keeping the others fixed to compare the ODMR fitting parameters of $\delta D$, the absolute error of $D$. According to which, the optimal combination of the experimental parameters was fixed. The Notice that the data of the same set of experiments were obtained at the same temperature. The comparison of different measurement parameters is showed in Fig. 7. Fig 7(a) shows the effects from count time and preparation time, and we selected 0.8 s as the count time because of higher accuracy. The effect of the preparation time in the pulse sequence is not significant so that 0.001 s was used. The microwave frequency step (Fig. 7(b)) is set to be 0.8 s. We selected 1.0 A as the output current of the laser, that is, 5.6 mW in power, on which the accuracy and the stability of data were slightly higher than those in 3.3 mW and the impact of the heating effect was not so significant as that on 8.1 mW (Fig. 7(c)). Although it was demonstrated that a greater microwave power led to a larger contrast of ODMR spectrum, we determined -5 dBm as the optimal microwave power in consideration of the durance of the copper wire and the heating effect of microwave (Fig. 7(d)).

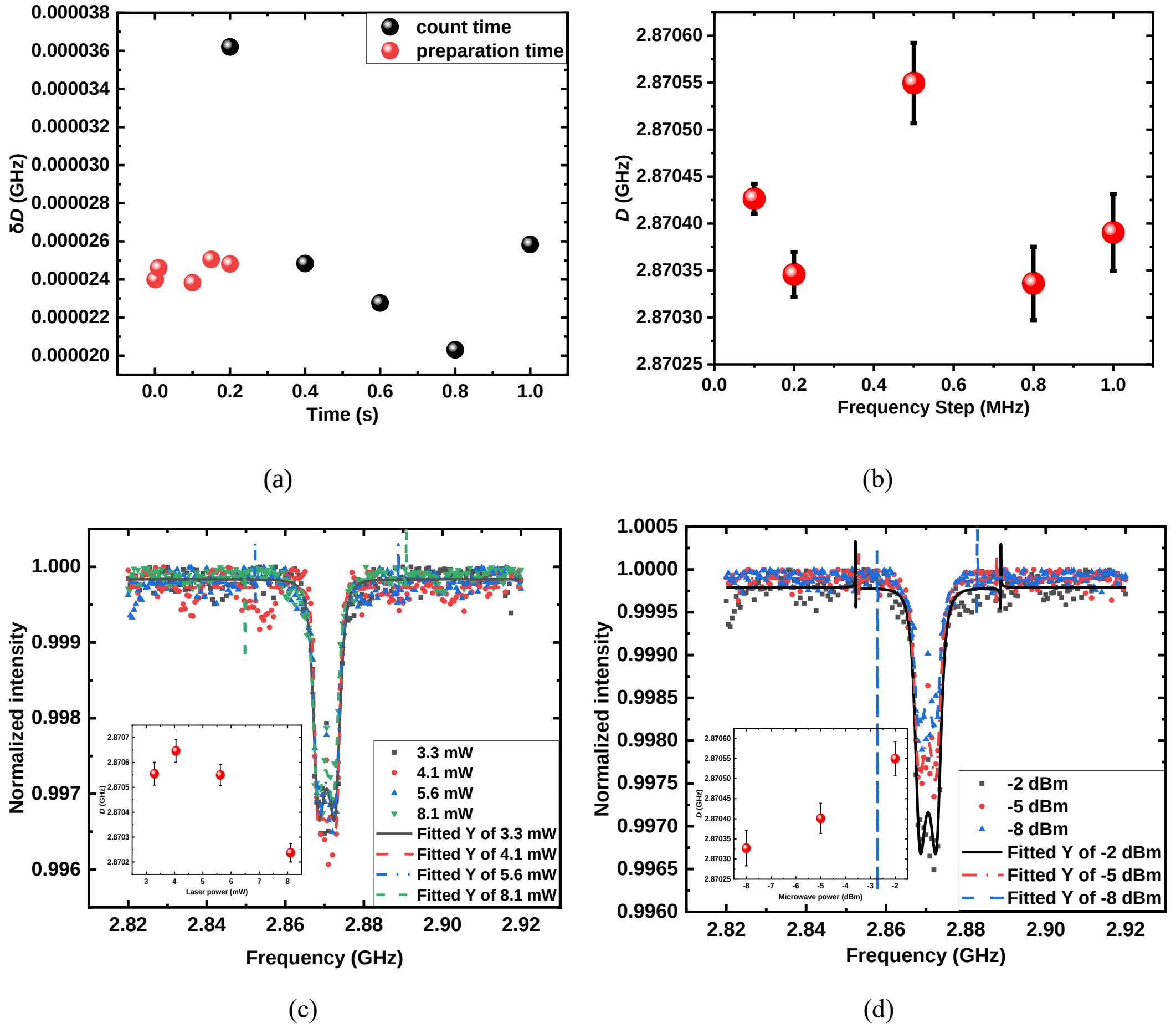

FIG. 7. The comparison of different measurement parameters. (a) $\delta D$ as a function of the preparation time and the count time. (b) The fitting results in the case of different microwave frequency step. (c) The graph of the fitting results in the case of different laser power. The inserted diagram shows $D$ and $\delta D$ as a function of the laser power. (d) The graphs of the spectra in the case of different microwave power. The inserted diagram shows $D$ and $\delta D$ as a function of the microwave power.

## C. Results of the relationship between *D* and *T*

We controlled the temperature from 298.15 K to 383.15 K by a step of 5 K and observed the ODMR spectrum with the optimal combination of parameter settings. At each temperature, three ODMR spectra were measured to understand the repeatability of the measurement. After that, we lowered the temperature from 383.15 K to 298.15 K by a step of 5 K and repeated the procedure.

We carried on two rounds of the heating and cooling process and calculated the average value of *D* obtained from the heating experiment and the cooling experiment at each temperature point, whose amount is 12 in total. The experimental data and the uncertainty are listed in Table 1 and showed in Fig. 8. As depicted in Fig. 8, the dip shifts towards a lower frequency as temperature rises, which is consistent with theoretical prediction.

TABLE 1. Experimental data and the uncertainty of *T* and *D*.

| $T$/K | $u(T)$/K | $D$/GHz | $u(D)$/GHz |
|---|---|---|---|
| 298.15 | 0.019 | 2.87007 | $5.82895 \times 10^{-5}$ |
| 303.15 | 0.022 | 2.86969 | $4.45660 \times 10^{-5}$ |
| 308.15 | 0.031 | 2.86928 | $4.77452 \times 10^{-5}$ |
| 313.15 | 0.040 | 2.86891 | $5.11100 \times 10^{-5}$ |
| 318.15 | 0.043 | 2.86850 | $4.70107 \times 10^{-5}$ |
| 323.15 | 0.062 | 2.86810 | $5.17925 \times 10^{-5}$ |
| 328.15 | 0.076 | 2.86766 | $3.62268 \times 10^{-5}$ |
| 333.15 | 0.092 | 2.86719 | $6.50587 \times 10^{-5}$ |
| 338.15 | 0.097 | 2.86675 | $5.43186 \times 10^{-5}$ |
| 343.15 | 0.098 | 2.86629 | $5.35958 \times 10^{-5}$ |
| 348.15 | 0.105 | 2.86582 | $3.93330 \times 10^{-5}$ |
| 353.15 | 0.134 | 2.86534 | $5.00841 \times 10^{-5}$ |
| 358.15 | 0.147 | 2.86485 | $7.27117 \times 10^{-5}$ |
| 363.15 | 0.157 | 2.86438 | $7.43089 \times 10^{-5}$ |
| 368.15 | 0.161 | 2.86385 | $7.80429 \times 10^{-5}$ |
| 373.15 | 0.259 | 2.86340 | $8.54464 \times 10^{-5}$ |
| 378.15 | 0.288 | 2.86289 | $7.17036 \times 10^{-5}$ |
| 383.15 | 0.311 | 2.86234 | $5.55988 \times 10^{-5}$ |

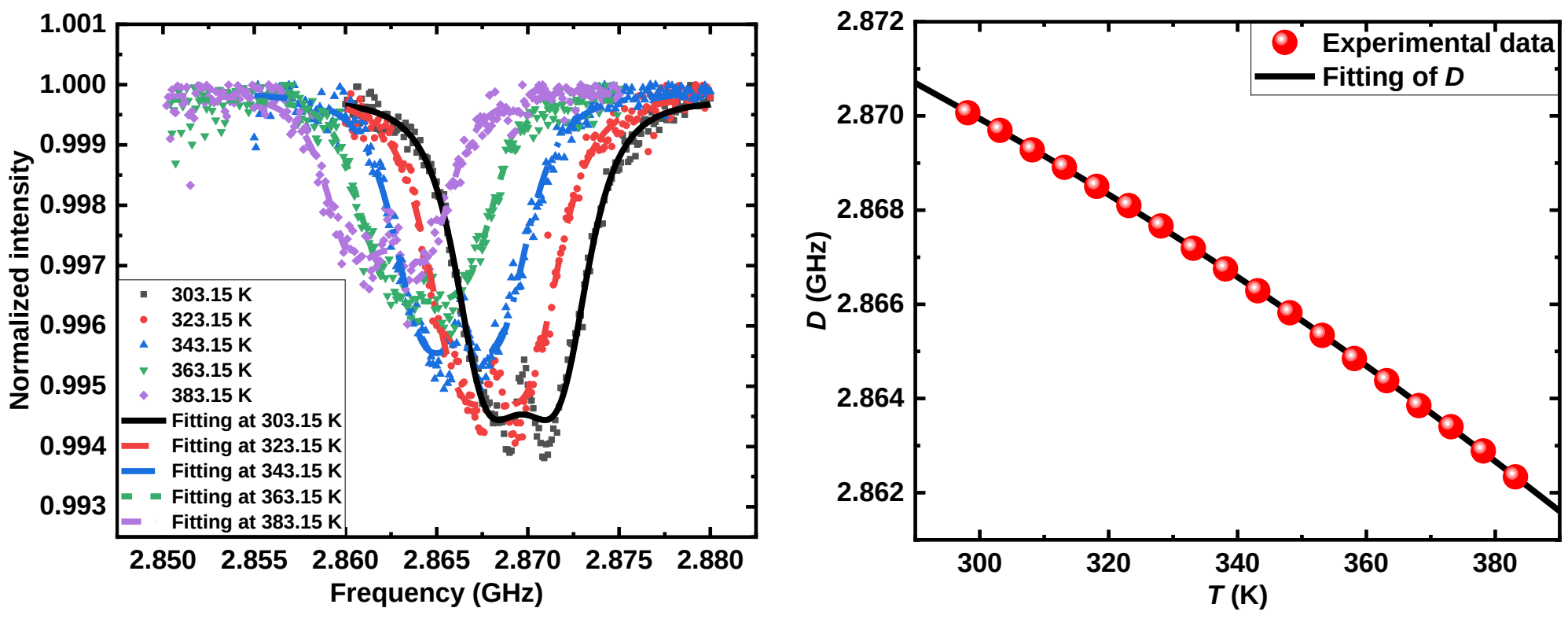

(a) (b)

FIG. 8. The graphs of ODMR spectrum lines and *D* at different temperature. (a) ODMR spectra and curve fittings at different temperatures. (b) The experimental and fitting of the average *D* and *T*.

The standard uncertainty budgets of *T* and *D* are listed in Table. 2. For the temperature measurement, the temperature inhomogeneity accounts for larger uncertainty of *T* in the lower temperature range while the fluctuation of controlling temperature becomes the dominant factor as temperature goes up. The uncertainty of *D* origins from four aspects: ODMR spectrum fitting, the repeatability of three measurements at one temperature, the reproducibility for four heating-cooling procedures, and microwave source. Except for microwave source, the other three all occupy substantially of the combined uncertainty. Finally, we figured out the uncertainty of *T* as 0.013 K ~0.311 K and the uncertainty of *D* as (3.62268~8.54464) × $10^{-5}$ GHz. Accordingly, the performance of *D*-*T* relationship calibration could be further improved by enhancing the stability of temperature control.

TABLE 2. Different determinations of *T* and *D* with their uncertainty budget and correlations.

| Uncertainty resource of *T* | $u(T)$/K | Uncertainty resource of *D* | $u(D)$/GHz |
|---|---|---|---|
| The uncertainty of the thermometer | 0.001 | Fitting | (2.49477~3.96674) × $10^{-5}$ |
| Pt-100 calibration | 0.0054~0.0065 | Repeatability | (1.59529~6.48305) × $10^{-5}$ |
| Temperature inhomogeneity | 0.0092~0.0279 | Reproducibility | (0~4.74561) × $10^{-5}$ |
| Temperature fluctuation | 0.0065~0.3092 | Frequency uncertainty of microwave source | $2.9 \times 10^{-5}$ |
| Combined uncertainty/K | 0.013~0.311 | Combined uncertainty/GHz | $(3.62268\text{~}8.54464) \times 10^{-5}$ |

The average *D* and *T* could be described accurately by a second-order polynomial with the *R*-square up to 0.99992 (Fig. 8(b)), expressed as

$$D = a_0 + a_1 T + a_2 T^2 \quad (4)$$

where $a_0$ = 2.87749(±0.00096) GHz, $a_1$ = 2.679(±0.568) × $10^{-5}$ GHz/K, $a_2$ = -1.7312(±0.0833) × $10^{-7}$ GHz/K$^2$. The residuals of *D* were plotted in Fig. 9 with a random distribution within ±0.00004 GHz.

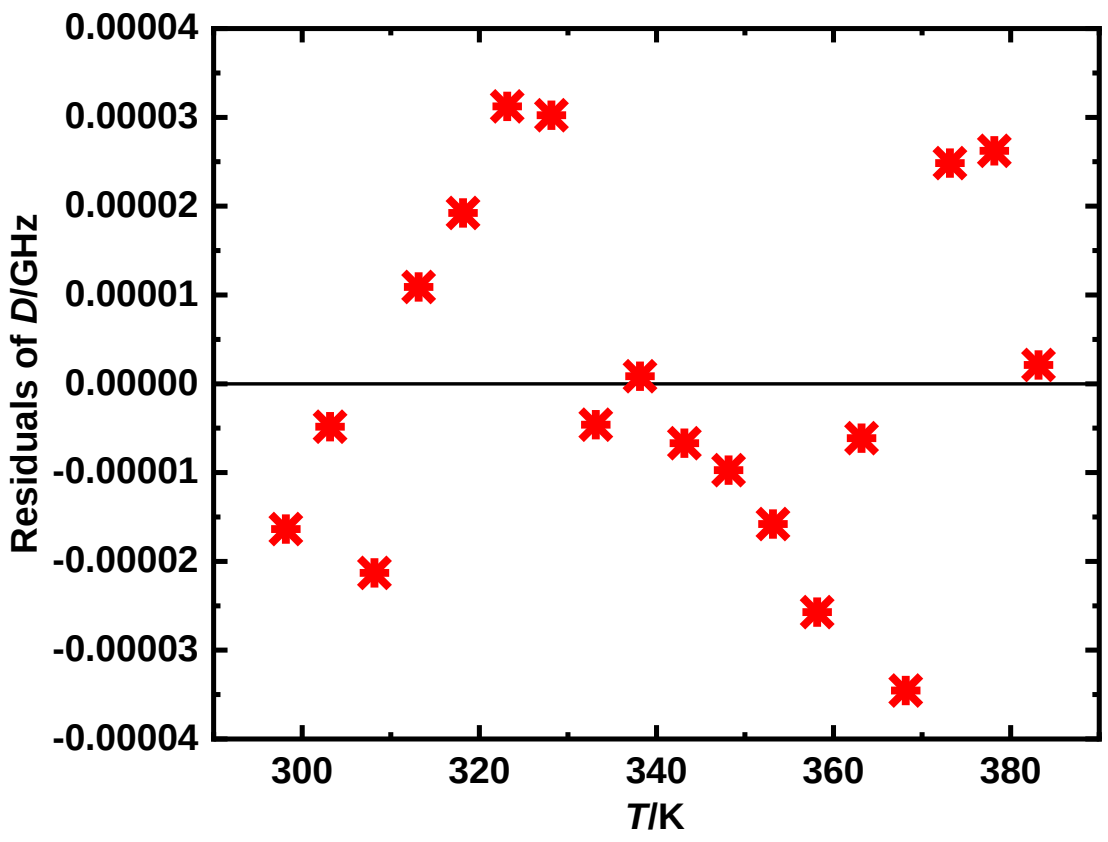

FIG. 9. Residuals of *D* in the second-order polynomial fitting.

In previous works, there are two typical relations between *D* and *T* covering the temperature range, that is,

$$D_1(T) = 2.8697 + 9.7 \times 10^{-5} T - 3.7 \times 10^{-7} T^2 + 1.7 \times 10^{-10} T^3, 300\ \text{K} < T < 700\ \text{K} \quad (5)$$

in Ref. [14], and

$$D_2(T) = 2.870 + 6 \times 10^{-5} T - 2.3 \times 10^{-7} T^2, 300\ \text{K} < T < 600\ \text{K} \quad (6)$$

in Ref. [9]. We plotted eqs. 4-7 in Fig. 10(a) and it is obvious that the relation obtained by us presents a high agreement with that in Ref. [14] but deviates from that in Ref. [9]. We also extrapolated our fitting equation down to 0 K and up to 700 K, and compared it with the work in Ref. [23] in the temperature range from 5.6 K to 295 K,

$$D_3(T) = 2.87771 - 4.625 \times 10^{-6}T + 1.067 \times 10^{-7}T^2 - 9.325 \times 10^{-10}T^3$$
$$+1.739 \times 10^{-12}T^4 - 1.838 \times 10^{-15}T^5 \quad (7)$$

The extrapolation of present *D*-*T* equation to high temperature agrees well with that in Ref. [14] and the extrapolation to low temperature agrees well with the relation obtained in Ref. [23], which proves the reliability of the experimental data.

The first order differential of the relationship between *D* and *T* from this work is

$$\frac{\mathrm{d}D}{\mathrm{d}T} = 2.679 \times 10^{-5} - 3.4624 \times 10^{-7}T \quad (8)$$

The comparison of d*D*/d*T* as a function of *T* in several references, including Ref. [14], Ref. [9], Ref. [24]. and Ref. [22] and this work is plotted in Fig. 10(b). To figure out the origin of consistency and deviation between different researches, we compared the experimental conditions of different works in Table 3. CW-ODMR was applied in all the references. The relationship of *D*-*T* in Ref. [14], Ref. [23] and this work, are highly consistent even in extrapolated range because of using similar samples. The difference between the trends of d*D*/d*T* origins from two aspects: the discrepancy between the manufacturing technology and NV concentration of the diamond samples, which can be explained by eq. 2. At the same temperature, the electron density $\eta$ increases with the increasing NV concentration, and the parameters involved in the formula, $\langle \frac{1}{r_{12}^3} - \frac{3z_{12}^2}{r_{12}^5} \rangle$, which are regarded as the internal attribution of NV centers microscopically, remained unchanged. Therefore, lower concentration of NV centers will lead to a smaller value of *D* at the same temperature, which is proved by the deviation of the result in Ref. [9] from the others in Fig. 10(a). When temperature varies, the distance between NV lattice changed so that the value of $\langle \frac{1}{r_{12}^3} - \frac{3z_{12}^2}{r_{12}^5} \rangle$ changes results in a dependence of *D* on *T*. The results from present work and Ref. [14] and Ref. [9] confirm that d*D*/d*T* varies with temperature. Therefore, it is necessary to calibrate the value of d*D*/d*T* at different temperatures to obtain reliable measurement results. Furthermore, for the application of the NV centers thermometry, it is necessary to check the samples' properties when using a reference's *D*-*T* relationship.

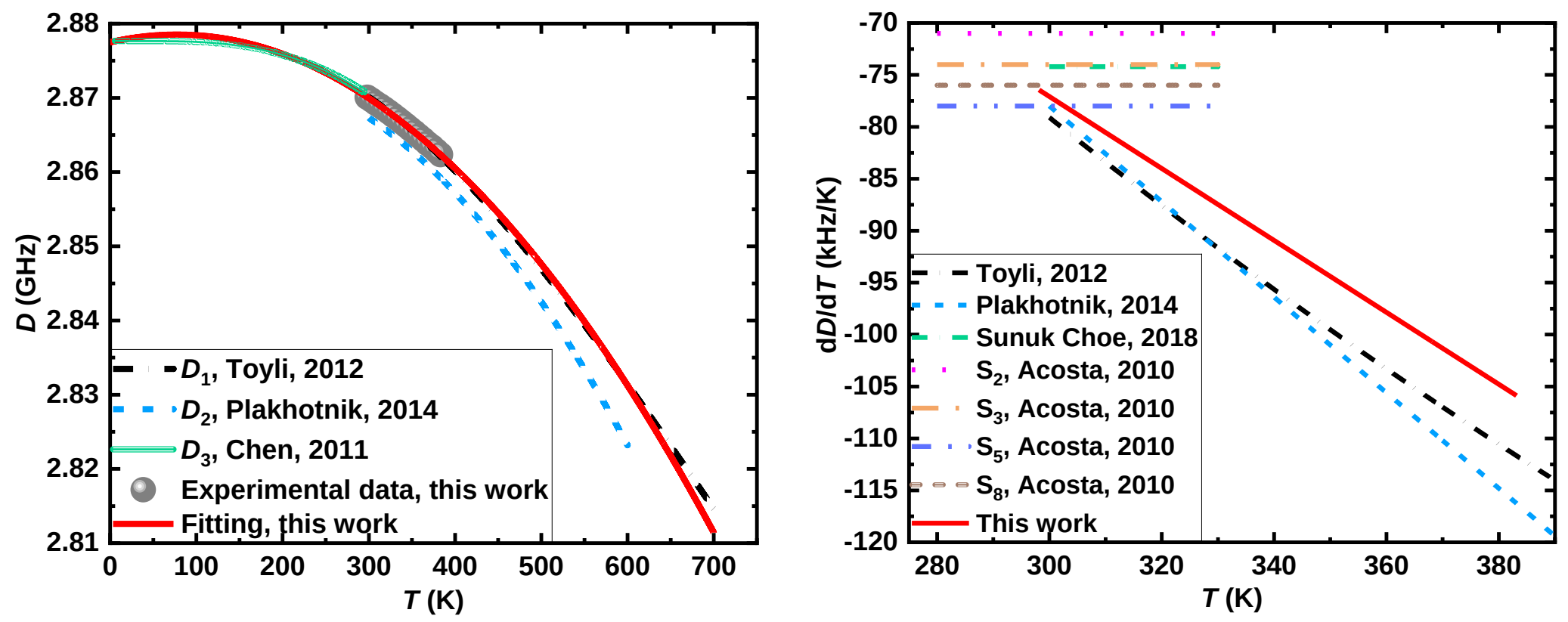

(a) (b)

FIG. 10. (a) Comparison of four *D*-*T* relationship curves obtained in Ref. [14], Ref. [9], Ref. [23], and this work. (b) Comparison of $\mathrm{d}D/\mathrm{d}T$ of the works from different references.

TABLE 3. Comparison of the technique details between different measurements.

| Reference | Diamond type | Manufacturer | Manufacturing technology | NV concentration |
|---|---|---|---|---|
| Ref. [14] | Single-crystal diamond | / | CVD (Chemical Vapor Deposition) | $[N_S^0] < 5$ ppb |
| Ref. [9] | Type Ib HPHT nano-diamond crystals | / | HPHT (High Pressure and High Temperature) | 15 $NV^-$ centers each |
| Ref. [24] | Single-crystal diamond | Element-Six | CVD | $[N_S^0] < 5$ ppb |
| Ref. [23] | Single-crystal diamond | Element-Six | CVD | [N]<1 ppm |
| Ref. [22] | Single-crystal diamond | / | $S_3$: CVD; $S_2$, $S_5$, $S_8$: HPHT | $S_2$: [$NV^-$]=16 ppm;<br>$S_3$: [$NV^-$]=10 ppb;<br>$S_5$: [$NV^-$]=12 ppm;<br>$S_8$: [$NV^-$]=0.3 ppm |
| This work | Single-crystal diamond | Element-Six | CVD | $[N_S^0] < 5$ ppb, [NV]<0.03 ppb |

# IV. CONCLUSION

To realize a reliable relationship between the zero-field splitting parameter *D* and temperature *T* for ensemble NV centers in diamond, we fabricated an optical fiber sensor coupled with NV centers, built up a CW-ODMR experimental system and a close thermostatic chamber for better thermal equilibrium. The measurement parameters and fitting function for the ODMR spectrum were compared and optimized. The ODMR experiments in the temperature range from 298.15 K to 383.15 K were carried out in four thermal circulates with a result of standard uncertainty of $u(D)$= (3.62268~8.54464) × $10^{-5}$ GHz and $u(T)$= (0.013~ 0.311) K. The experimental *D*-*T* was fitted using a second-order polynomial correlation with a maximal residual of 0.00004 GHz. We extrapolated the relationship of present *D*-*T* to a wider temperature range from 0 K to 700 K and it agrees well with other measured values. The first order differential of *D*-*T* also shows a temperature dependence, which means it is necessary to calibrate the value of d*D*/d*T* at different temperatures for accurate measurement. The diamond sample's properties result in a different *D*-*T* relationship mainly because the NV concentration results in different electron density and the manufacturing procedure results in different thermal expansion. In the next future, we will continue further research especially from the metrological point of view to develop NV centers as a practical and accurate micro-nano scale thermometry.

## ACKNOWLEDGEMENT

This work was supported by the Fundamental Research Program of National Institute of Metrology, China (No. AKYZD1904-2) and China Postdoctoral Science Foundation (No. 2021M703049). The authors greatly appreciate the helpful discussion from Dr. Mark Plimmer.